\documentclass[final]{IEEEtran}
\usepackage{amsthm,amssymb,graphicx,multirow,amsmath,color,amsfonts}
\usepackage[update,prepend]{epstopdf}
\usepackage[noadjust]{cite}
\usepackage{tikz}
\usepackage{bbm} 
\usepackage{pdfpages}
\usepackage{tabulary}
\usepackage{multirow}
\usepackage{comment}
\usepackage{subfigure}
\usepackage{float}

\def\nb0{{\mathbf{0}}}
\def\nb1{{\mathbf{1}}}

\newtheorem{lemma}{Lemma}

\newtheorem{definition}{Definition}

\newtheorem{theorem}{Theorem}

\newtheorem{remark}{Remark}

\allowdisplaybreaks 
\begin{document}
	\graphicspath{{./Figures/}}
	\title{
		Performance Evaluation of UAV-enabled Cellular Networks with Battery-limited Drones}
	\author{ Yujie Qin, Mustafa A. Kishk, {\em Member, IEEE}, and Mohamed-Slim Alouini, {\em  Fellow, IEEE} 
		\thanks{Yujie Qin is with University of Electronic Science and Technology of China (UESTC), Chengdu 610054, China. Mustafa A. Kishk and Mohamed-Slim Alouni are with King Abdullah University of Science and Technology (KAUST), Thuwal 23955-6900, Saudi Arabia. This work was done during Yujie Qin's internship in KAUST.  (e-mails: yujie.qin@kaust.edu.sa; mustafa.kishk@kaust.edu.sa; slim.alouini@kaust.edu.sa)} 
	}

	\maketitle
	
	\begin{abstract}
		Unmanned aerial vehicles (UAVs) can be used as flying base stations (BSs) to offload Macro-BSs in hotspots. However, due to the limited battery on-board, UAVs can typically stay in operation for less than 1.5 hours. Afterward, the UAV has to fly back to a dedicated charging station that recharges/replaces the UAV's battery. In this paper, we study the performance of a UAV-enabled cellular network while capturing the influence of the spatial distribution of the charging stations. In particular, we use tools from stochastic geometry to derive the coverage probability of a UAV-enabled cellular network as a function of the battery size, the density of the charging stations, and the time required for recharging/replacing the battery.
	\end{abstract}
	
	\section{Introduction} \label{sec:intro}
	
	Owing to their freedom of mobility and relocation flexibility, UAVs can be used as flying BSs to provide cellular coverage~\cite{8713514}. In addition, the deployment at relatively high altitudes, compared to terrestrial BSs (TBSs), increases the chances of establishing a line-of-sight (LoS) link with the ground users. This has motivated many works in recent years to study and analyze the performance of UAV-enabled cellular networks~\cite{8660516}.
	
	However, the feasibility of UAV-enabled cellular networks still faces many challenges such as limited energy resources, which leads to limited flight time~\cite{DBLP:journals/corr/abs-1907-04299}. This, in turn, forces the UAV to interrupt its operation as a flying BS on a regular basis in order to fly back to the charging station to recharge/swap its battery. During charging, the users in the UAV's coverage area need to rely on other resources for cellular services, such as other nearby TBSs~\cite{8833522}.
	
	Multiple solutions have been provided in the literature to overcome this issue. Authors in~\cite{8648453} proposed UAV swapping, which suggests having an up and ready UAV to replace the operating UAV as soon as its battery gets drained. Another less expensive solution is battery swapping, which is based on replacing the drained battery with a fully charged one, instead of waiting for the drained battery to be recharged. Authors in ~\cite{8866716} studied the performance of a laser-powered UAV system, where laser beams are used to wirelessly charge the UAV's battery while it is hovering and providing cellular service. Authors in~\cite{DBLP:journals/corr/abs-1907-04299,kishkmag2019,bushnaq2020cellular} proposed using tethered UAVs, where a trade-off arises between having a stable power supply through the tether and limiting the mobility of the UAV. Authors in~\cite{7918510} proposed an optimal UAV placement approach that maximizes the coverage area while reducing its transmit power. However, given that propulsion power consumption of the UAV dominates the power consumed for wireless communication, solutions that rely on enhancing the communication energy efficiency are not expected to dramatically increase the UAV flight time. 
	
	
	\textit{Contributions}.
	This paper studies the influence of the UAV's limited battery, the density of the charging stations, and the recharging/swapping time on the performance of the UAV-enabled cellular network. In particular, we use tools from stochastic geometry to derive the availability probability of the UAV. Next, we use this result to study the impact of the availability probability of the UAV on the coverage probability of the cellular network.
	\section{System Model} \label{sec:SysMod}
	We consider a UAV-enabled cellular network composed of TBSs and UAVs, where the UAVs are located at the centers of hotspots. In order to model the locations of the users in the hotspot, one of the most popular models in literature is Poisson cluster process (PCP)~\cite{7809177}. There are two possible types of PCP: (i) Thomas cluster process and (ii) Matern cluster process (MCP). In this paper, we model the locations of the users in hotspots using MCP. In particular, the hotspots are modeled as randomly located disks with fixed radius $r_{\rm c}$. The centers of the disks, above which the UAVs are deployed, are modeled as a Poisson point process (PPP). Within each disk, the users are uniformly distributed. The UAVs are assumed to hover at a fixed altitude $h$ above each hotspot center. The locations of the TBSs are modeled as a PPP $\Phi_{\rm TBS}$ with density $\lambda_{\rm T}$.
	
	Unlike existing literature, the main objective of this paper is to study the impact of the spatial distribution of the charging stations on the performance of the above setup. We model the locations of charging stations as a PPP $\Phi_{\rm c}$ with density $\lambda_{\rm c}$.
	\subsection{UAV's Availability}
	We consider a scenario where each UAV is supposed to fly back to its nearest charging station before running out of energy. During traveling to/from the charging station, as well as during recharging, the UAV is considered unavailable and can not provide service. 
	
	\begin{definition}[Availability probability] \label{def:1}
		We define the event $\mathcal{A}$ that indicates the availability of the UAV. 
		Conditioned on the distance between the hotspot center and the nearest charging station $R_{\rm s}$, the availability probability of the UAV is
		\begin{align}\label{eq:pa_rs}
		P_{(\rm a|R_{\rm s})}&=\mathbb{P}(\rm \mathcal{A} | R_{\rm s}) \nonumber\\
		&=\frac{T_{\rm se}}{T_{\rm se}+T_{\rm ch}+T_{\rm tra}},
		\end{align}
		where $T_{\rm tra}$ is the required time to travel to and from the nearest charging station, $T_{\rm se}$ denotes the time spent at the hotspot center to provide cellular service, and ${T}_{\rm ch}$ presents the total time required for recharging or swapping. 
	\end{definition}
	Each of $T_{\rm tra}$ and $T_{\rm se}$ can be formally defined as follows:
	\begin{align}\label{eq:T}
	T_{\rm se}&=\frac{B_{\rm max}-2P_{\rm m}\frac{R_{\rm s}}{V}}{P_{\rm s}}, \nonumber\\
	T_{\rm tra}&=\frac{2R_{\rm s}}{V},
	\end{align}
	where ${B}_{\rm max}$ is the UAV battery size, $P_{\rm m}$ denotes the power consumption during traveling, $V$ is the UAV's velocity during traveling, and $P_{\rm s}$ is the power consumption during hovering at the hotspot center, which includes both the propulsion power and the total communication power. Note that for power consumption during traveling, we focus on the power consumed to travel the horizontal distance $R_{\rm s}$ since it is typically larger than the power consumed during landing, after reaching the charging station. 
	
	
	Now, the availability probability of the UAV can be derived by taking the expectation of the conditioned probability provided in Definition~\ref{def:1} as follows
	\begin{align}
	P_{\rm a}=\mathbb{E}_{\Phi_{\rm c}}\left[\frac{T_{\rm se}}{T_{\rm se}+T_{\rm ch}+T_{\rm tra}}\right].
	\end{align}
	
	\subsection{Power Consumption}
	The value of ${P}_{\rm s}$ is assumed to be fixed, while ${P}_{\rm m}$ is given as follows \cite{8663615}
	\begin{align}
	P_{\rm m}=P_{\rm 0}\left(1+\frac{3V^2}{U_{\rm tip}^{2}}\right)+\frac{P_{\rm i}v_{\rm 0}}{V}+\frac{1}{2}d_{\rm 0}\rho s AV^{3},
	\end{align}
	where $U_{\rm tip}^{2}$ is the top speed of the rotor blade, $v_{\rm 0}$ is the mean rotor induced velocity in hover, $\rho$ is the air density, $A$ is the rotor disc area, $d_{\rm 0}$ is fuselage drag ratio, $V$ is the velocity of the UAV, and $P_{0}$ and $P_{i}$ represent the UAV's blade profile power and induced power in hovering status, respectively. (See (12) and (64) in Ref \cite{8663615} for more details.) Consequently, the energy consumed during traveling to or from the charging station is 
	\begin{align}
	\label{1}
	E_T&=\frac{R_{\rm s}}{V}P_{\rm m} \nonumber\\
	&=\frac{R_{\rm s}}{V}\left(P_{\rm 0}\left(1+\frac{3V^2}{U_{\rm tip}^{2}}\right)+\frac{P_{\rm i}v_{\rm 0}}{V}+\frac{1}{2}d_{\rm 0}\rho s AV^{3}\right).
	\end{align}
	In the numerical results section, we use the value of $V$ that minimizes $E_T$, referred to as $V_{\rm opt}$.

	\subsection{User Association}
	We assume that each user connects to the UAV deployed at its hotspot center if it is available. Otherwise, the user connects to the nearest TBS. Throughout this paper, we focus our analysis on a randomly selected user inside the hotspot, which is referred to as the reference user.
	
	When the user associates with the UAV, the average received power is
	\begin{align}\label{7}
	p_{\rm u}&=\left\{ 
	\begin{aligned}
	p_{\rm L}=\eta_{\rm L}\rho_{\rm u}G_{L}R_{\rm U}^{-\alpha_{\rm L}},  & \quad \text{\rm in case of LoS},\\
	p_{\rm N}=\eta_{\rm N}\rho_{\rm u}G_{N}R_{\rm U}^{-\alpha_{\rm N}},  & \quad \text{\rm in case of NLoS},\\
	\end{aligned} \right.
	\end{align}
	where $\rho_u$ is the transmission power of the UAV, $R_{\rm U}$ is the distance between the reference user and the UAV, $\alpha_{\rm L}$ and $\alpha_{\rm N}$ present the path-loss exponent, $G_{L}$ and $G_{N}$ are the fading gains that follow gamma distribution with shape and scale parameters $(m_{\rm L},\frac{1}{m_{\rm L}})$ and $(m_{\rm N},\frac{1}{m_{\rm N}})$, $\eta_{\rm L}$ and $\eta_{\rm N}$ denote the mean additional losses for LoS and NLoS transmissions, respectively.
	
	According to ~\cite{6863654}, the probability that the  UAV has a LoS channel to the reference user is given as
	
	\begin{align}
	P_{\rm L}(R_{\rm U}) & =  \frac{1}{1+a \exp(-b(\frac{180}{\pi}\arctan(\frac{h}{\sqrt{R_{\rm U}^2-h^2}})-a))} ,
	\end{align}
	where $a$ and $b$ are constants that related to the environment and $h$ is the altitude of the UAV. Moreover, the probability of NLoS is $P_{\rm N}(R_{\rm U})=1-P_{\rm L}(R_{\rm U})$.
	
	
	
	When the UAV is unavailable, the user associates with the nearest TBS. In that case,
	the received power $p_{\rm t}$ is given by
	\begin{align}
	\label{9}
	p_{\rm t}=\rho_{\rm t} H R_{\rm T}^{-\alpha_{\rm T}},
	\end{align}
	where $\rho_t$ is the transmission power of the TBS, $R_{\rm T}$ presents the distance between the reference user and the nearest TBS, and $H$ is the fading gain that follows exponential distribution with average power of unity.
	\begin{definition}[Coverage probability] \label{def:2} The total coverage probability conditioned on $R_{\rm s}$ is defined as 
		\begin{align}
		\label{11}
		&P_{\rm cov|R_{\rm s}} = P_{\rm (a|R_s)} P_{\rm cov,{U}}+(1-P_{\rm (a|R_s)})P_{\rm cov,{T}},
		\end{align}
		where $P_{\rm (a|R_s)}$ is  given in (\ref{eq:pa_rs}). The unconditioned coverage probability is given by
		\begin{align}
		\label{111}
		&P_{\rm cov} = P_{\rm a} P_{\rm cov,{U}}+(1-P_{\rm a})P_{\rm cov,{T}},
		\end{align}
		in which,
		\begin{align}
		\label{8}
		P_{\rm cov,{\{U,T\}}} = \mathbb{P}\left(\frac{p_{\{u,t\}}}{\sigma^2}\geq\beta\right) ,
		\end{align}
		where $\sigma^2$ is the noise power and $\beta$ is the signal-to-noise-ratio (SNR) threshold.
	\end{definition}
	
	\section{Performance Analysis}
	In this section, we provide the main results in this paper. We first derive the availability probability conditioned on the distance to the nearest charging station $R_s$, as well as the unconditioned availability probability. Finally, we use these results to study the coverage probability in the considered system setup.
	
	
	
	
	
	
	
	\subsection{Availability Probability}
	
	In this subsection, we analyze the statistics of the availability probability of the UAV.
	This analysis will be used to study the coverage probability in the next subsection.
	
	\begin{lemma}[Conditional Availability Probability]\label{lem:1}
		Given the value of $R_{\rm S}$, the availability probability is given by
		\begin{align}
		P_{(\rm a|R_{\rm s})}=\frac{{B}_{\rm max}V-2{P}_{\rm m}{R}_{\rm s}}{{B}_{\rm max}V-2{P}_{\rm m}{R}_{\rm s}+T_{\rm ch}P_{\rm s}V+2{R}_{\rm s}P_{\rm s}}.
		\end{align}
		\begin{IEEEproof}
			The above result follows directly by substituting for (\ref{eq:T}) in (\ref{eq:pa_rs}).
		\end{IEEEproof}
	\end{lemma}
	\begin{remark}
		Note that the above expression only holds if $R_{\rm s}\leq \frac{V B_{\rm max}}{2P_{\rm m}}$. Otherwise, $P_{(\rm a|R_{\rm s})}=0$. If this condition is not satisfied, the battery size is not large enough to support energy for the UAV to travel to and from the charging station. Hence, there will not be enough power for the UAV to serve the users in the hotspot. In addition, when $R_{\rm s}=0$, the maximum availability probability is achieved. In that case, $P_{(\rm a|R_{\rm s}=0)}=\frac{B_{\rm max}}{B_{\rm max}+P_{\rm s}T_{\rm ch}}$.
	\end{remark}
	Given that the value of $R_{\rm s}$ varies from one hotspot to the other,  in the below lemma, we derive the CDF of the conditional availability probability.
	\begin{lemma}[CDF of Conditional Availability Probability]\label{lem:2}
		The CDF of the conditional availability probability is given by
		\begin{align}
		\label{4}
		F_{P_{(\rm a|R_{\rm s})}}(x) = e^{-\lambda_{\rm c} \pi C(x)^2},
		\end{align}
		in which,
		\begin{align}
		C(x)&=\frac{V (B_{\rm max} (x-1)+P_{\rm s} T_{\rm ch} x)}{2 (P_{\rm m} (x-1)-P_{\rm s} x)},\\ 
		&0\leq x\leq\frac{B_{\rm max}}{P_{\rm s}T_{\rm ch}+B_{\rm max}} \label{6}.
		\end{align}
		\begin{IEEEproof}
			See Appendix~\ref{app:2}.
		\end{IEEEproof}
	\end{lemma}
	
	In the following theorem, we derive the availability probability.
	
	
	\begin{theorem}[Availability Probability]\label{lem:3}
		The availability probability of the UAV $P_{\rm a}$ is
		\begin{align}
		P_{\rm a}=\int_{0}^{\frac{B_{\rm max}}{P_{\rm s}T_{\rm ch}+B_{\rm max}}}1-e^{-\lambda_{\rm c} \pi C(x)^2}{\rm d}x.
		\label{10}
		\end{align}
		\begin{IEEEproof}
			The above expression follows by substituting the results in Lemma~\ref{lem:2} into
			\begin{align}
			P_{\rm a}=\mathbb{E}_{\Phi_{\rm c}}[P_{(\rm a|R_{\rm s})}]=\int_0^{\infty} 1-F_{P_{(\rm a|R_{\rm s})}}(x) {\rm d}x.
			\end{align}
		\end{IEEEproof}
	\end{theorem}
	
	\subsection{Coverage Probability}
	Using the results provided in the previous subsection, we can now study the coverage probability as explained in Definition~\ref{def:2}. First, we need to provide expressions for each of $P_{\rm cov,U}$ and $P_{\rm cov,T}$, which are provided next.
	
	\begin{lemma}[Coverage Probability]\label{lem:4} The coverage probability when associating with a UAV or a TBS are  
		\begin{small}
			\begin{align}
			&P_{\rm cov,U}=\sum_{k=0}^{m_{\rm L}-1}\frac{2}{r_{\rm c}^2k!}\int^{\sqrt{(h^2+r_c^2)}}_{h}P_{\rm L}(r)re^{-m_{\rm L}g_{l}(r)}(m_{\rm L}g_{l}(r))^{k}{\rm d}r \nonumber\\
			&+\sum_{k=0}^{m_{\rm N}-1}\frac{2}{r_{\rm c}^2k!}\int^{\sqrt{(h^2+r_c^2)}}_{h}P_{\rm N}(r)re^{-m_{\rm N}g_{n}(r)}(m_{\rm N}g_{n}(r))^{k}{\rm d}r,\\
			&P_{\rm cov,T} =\int^{\infty}_{0}2\pi r \lambda_{\rm T} e^{-\pi \lambda_{\rm T} r^{2}}e^{-g_{m}(r)}{\rm d}r,
			\end{align}
		\end{small}
		in which,
		\begin{align}
		g_{l}(r)&=\frac{\beta \sigma^2}{\eta_{\rm L} r^{-\alpha_{\rm L}}\rho_{\rm u}}, \nonumber\\
		g_{n}(r)&=\frac{\beta \sigma^2}{\eta_{\rm N} r^{-\alpha_{\rm N}}\rho_{\rm u}}, \nonumber\\
		g_{m}(r)&=\frac{\beta \sigma^2}{ r^{-\alpha_{\rm T}}\rho_{\rm t}}. \nonumber
		\end{align}
	\end{lemma}
	\begin{IEEEproof}
		See Appendix~\ref{app:3}.
	\end{IEEEproof}
	
	Due to the direct influence of the availability probability on the coverage probability, and the fact that $R_s$ varies from one cluster to the other, it is important to understand how the value of $R_s$ impacts the coverage probability. In the below lemma, we derive the CDF of the conditional coverage probability described in Definition~\ref{def:2}.
	
	\begin{lemma}[CCDF of Coverage Probability]\label{lem:5}
		The complementary cumulative distribution of coverage probability given $R_{\rm s}$ is 
		\begin{align}\label{5}
		F_{P_{\rm cov|R_{\rm s}}}(\theta)=1-F_{P_{(\rm a|R_{\rm s})}}\left(\frac{\theta-P_{\rm cov,T}}{P_{\rm cov,U}-P_{\rm cov,T}}\right),
		\end{align}
		where 
		\begin{align}
		P_{\rm cov,T}\leq\theta\leq \frac{B_{\rm max}({P_{\rm cov,U}-P_{\rm cov,T}})}{P_{\rm s}T_{\rm ch}+B_{\rm max}}+P_{\rm cov,T},
		\end{align}
		and $F_{P_{(\rm a|R_{\rm s})}}(x)$ is given in (\ref{4}).
		\begin{IEEEproof}
			The above result follows directly by substituting for $P_{\rm cov|R_{\rm s}}$ in (\ref{4}).
		\end{IEEEproof}
	\end{lemma}
	
	\begin{remark}\label{rem:cov}
		Note that the range of values of $\theta$ in the above result reflects the maximum and minimum achievable values of $P_{\rm cov|R_s}$. In particular, the minimum achievable value is $P_{\rm cov|R_s}=P_{\rm cov,T}$ reflects the scenario where $R_{\rm s}$ is too large that the UAV is always unavailable. On the other hand, the maximum achievable value is $P_{\rm cov|R_s}=\frac{B_{\rm max}({P_{\rm cov,U}-P_{\rm cov,T}})}{P_{\rm s}T_{\rm ch}+B_{\rm max}}+P_{\rm cov,T}$ reflects the scenario where $R_{\rm s}=0$.
		
		
	\end{remark}
	\begin{figure*}[htb]
		\centering
		\subfigure[]{\includegraphics[width=1\columnwidth]{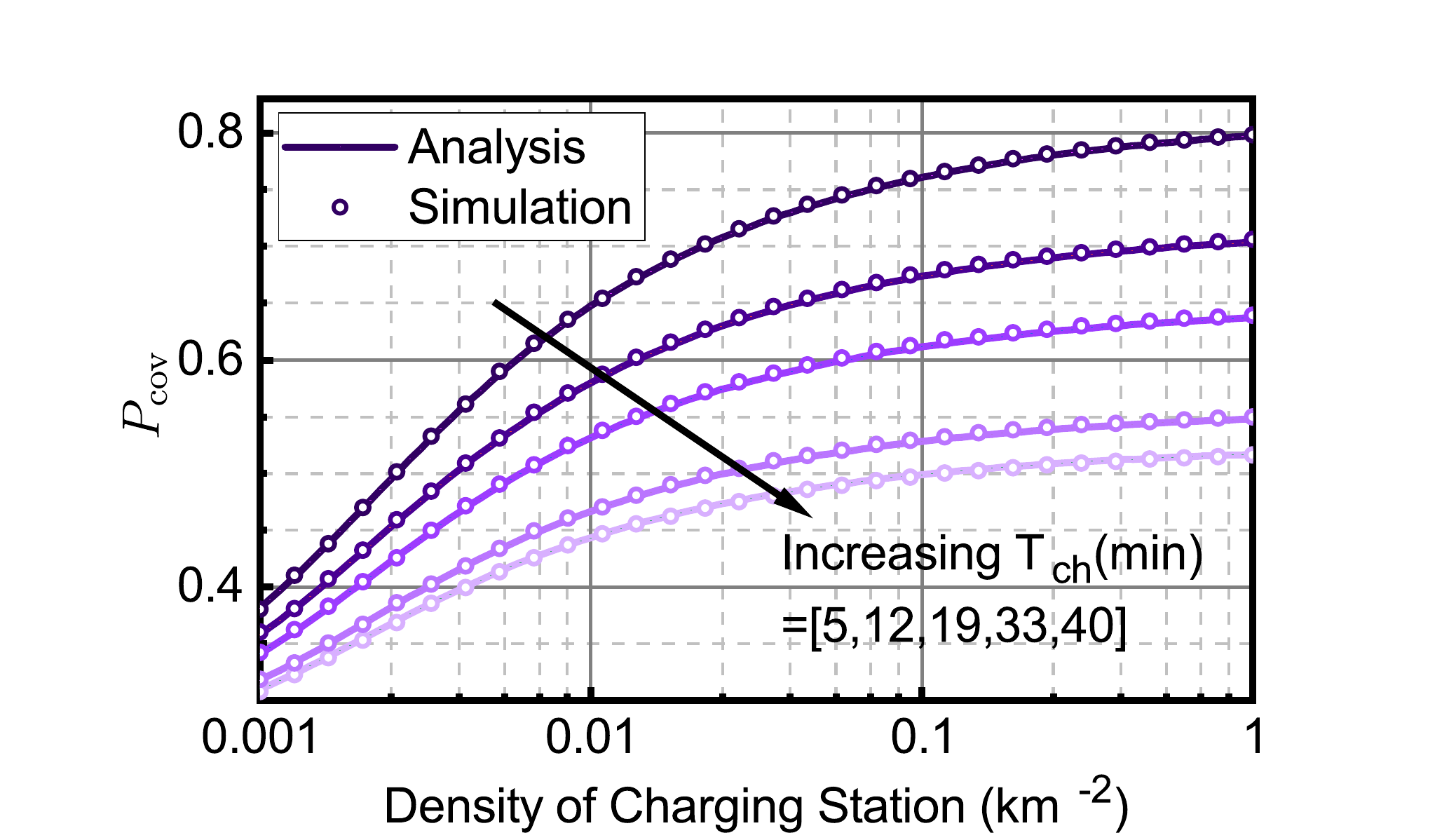}}
		\subfigure[]{\includegraphics[width=1\columnwidth]{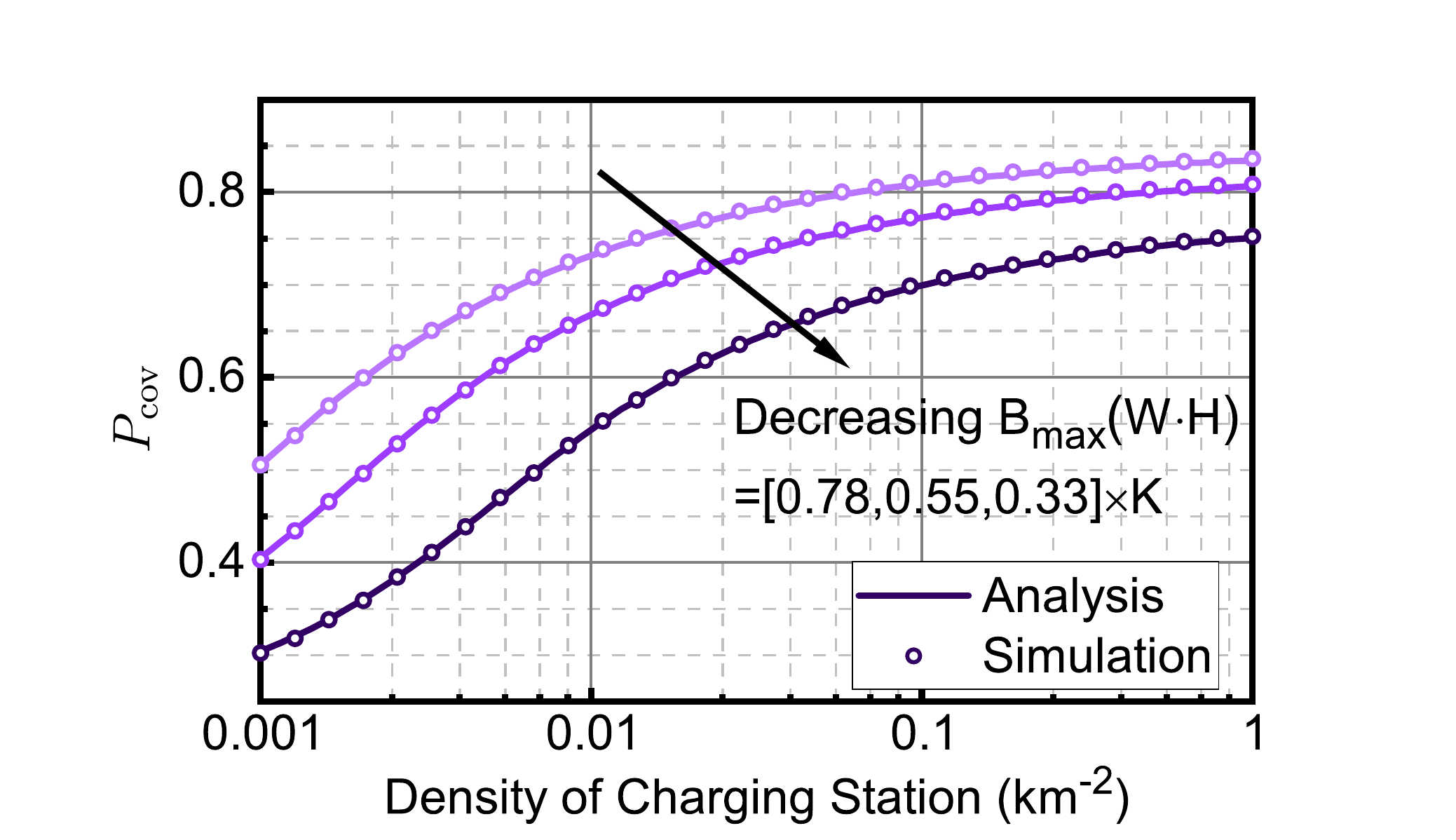}}
		\caption{The variation of the coverage probability with $\lambda_c$ at (a) different values of $T_{\rm ch}$, and (b) different values of $B_{\rm max}$ with $K=177.6$.}
		\label{fig:2}
	\end{figure*}
	\begin{figure*}
		\centering
		\subfigure[]{\includegraphics[width=1\columnwidth]{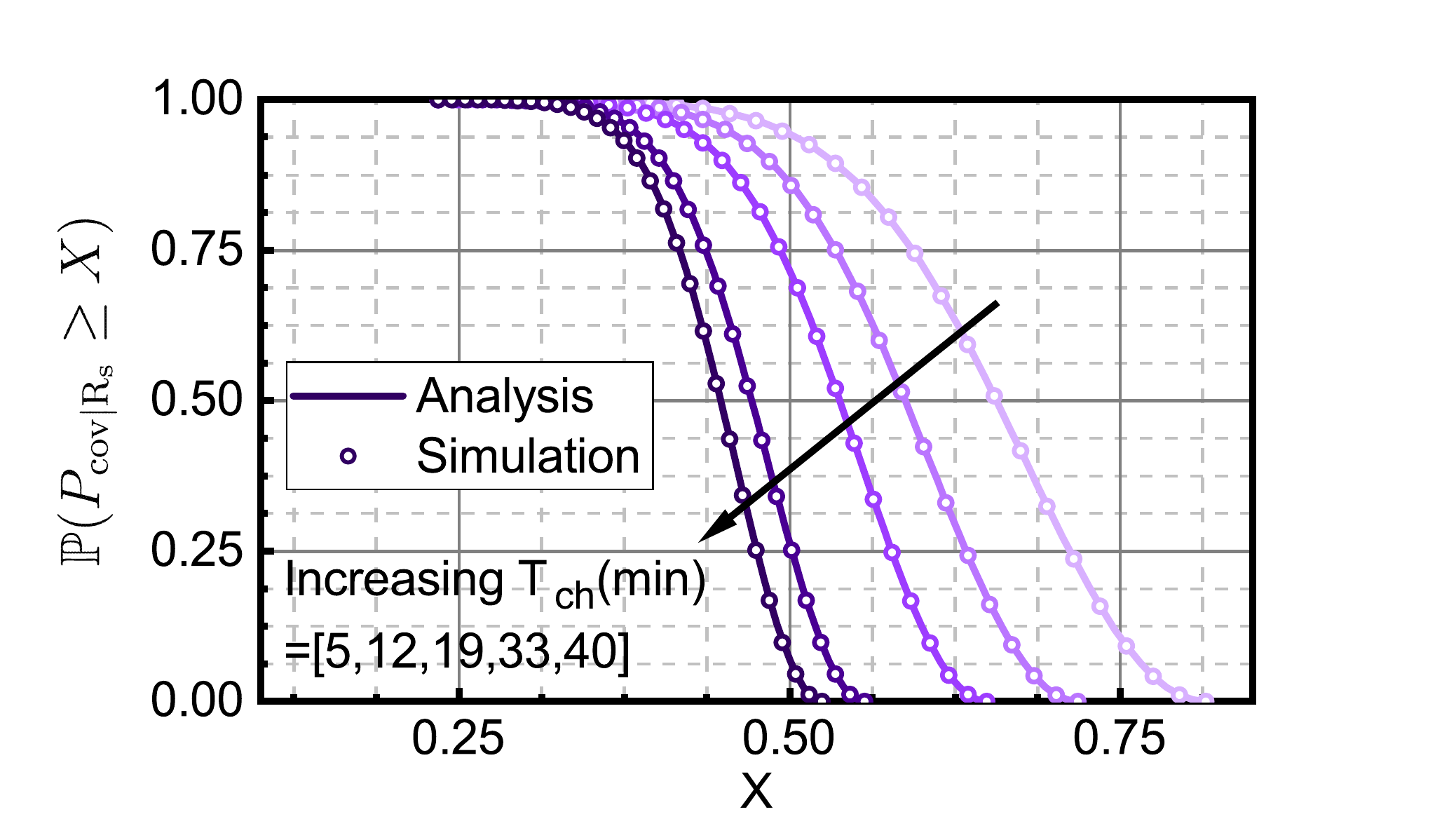}}
		\subfigure[]{\includegraphics[width=1\columnwidth]{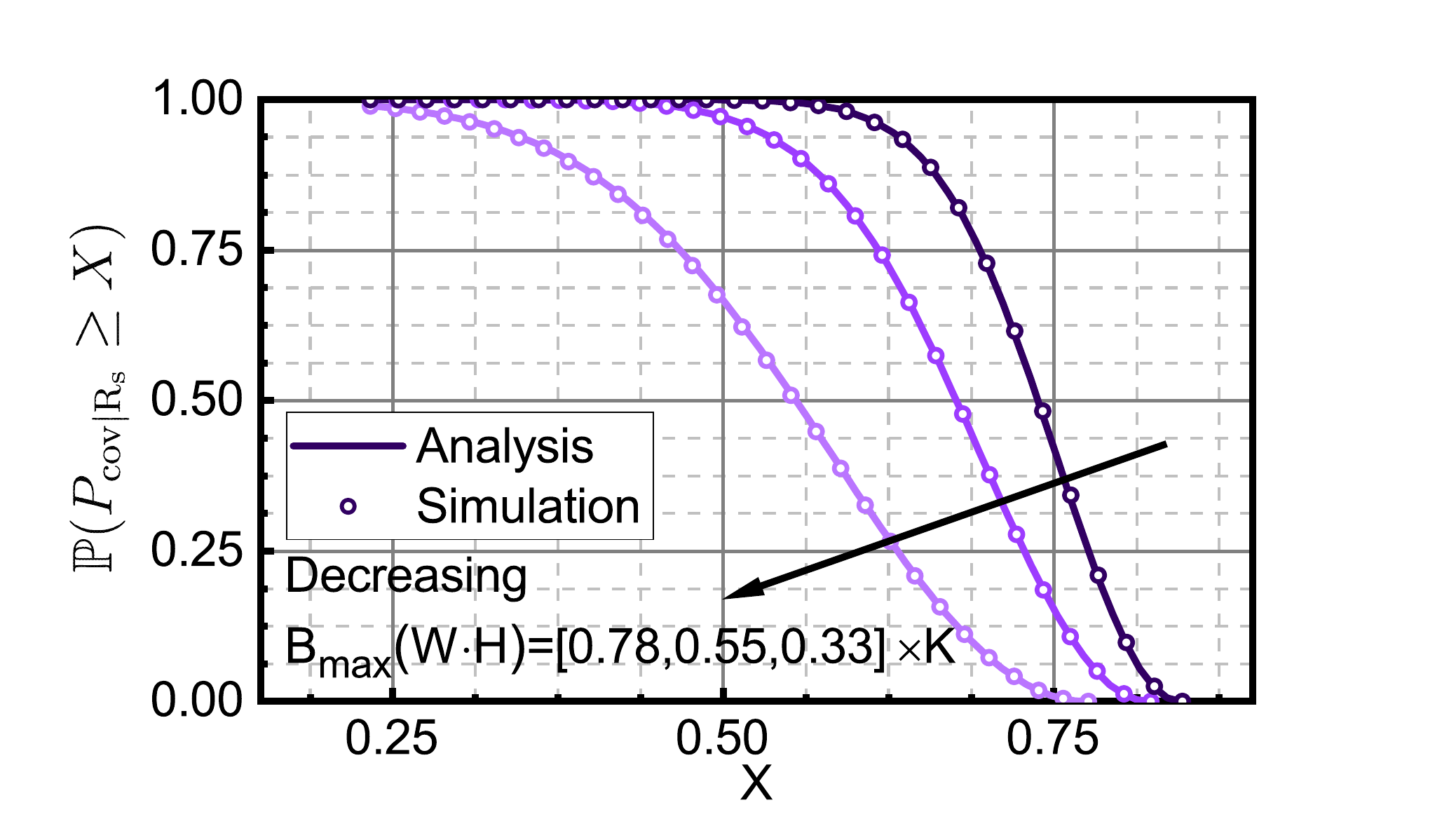}}
		\caption{The CCDF of $P_{\rm cov|R_s}$ at (a) different values of $T_{\rm ch}$, and (b) different values of $B_{\rm max}$ with $K=177.6$. }
		\label{fig:3}
	\end{figure*}
	\section{Numerical Results}
	
	In this section, we compute the value of the coverage probability using the derived analytical results and Monte-Carlo simulations to study the effect of the density of charging stations, the battery size, and the charging time. Unless otherwise specified, the values of the simulation parameters are summarized in Table \ref{table2}. 

	\begin{table}
		\caption{Simulation Parameters}
		\label{table2}
		\centering
		\begin{tabular}{cccc}
			\hline
			Power Consumption \cite{8663615}\\
			\hline
			$T_{\rm ch}$ & $5$ minutes & $B_{\rm max}$ & $88.8$ W$\cdot$H\\
			$P_{\rm s}$ & $177.5$ W & $\lambda_c$ & $10^{-2}$ km$^{-2}$\\
			$V_{\rm opt}$ & 18.46 m/s & $P_{\rm m}$& 161.8 W\\
			\hline
			Network Parameters\cite{8713514} &  &  & \\
			\hline    
			$\rho_{\rm u}$ &$0.1$ W &$\rho_{\rm t}$ & $10$ W\\
			$h$ & $60$ m & $\sigma^{2}$ & $10^{-9}$ W\\
			$r_{\rm c}$ & $100$ m & $\beta$ & $20$ dB  \\
			\hline
			LoS Parameters & & NLoS Parameters &  \\
			\hline
			$\eta_{\rm L}$ & $0$ dB & $\eta_{\rm N}$ & $20$ dB\\
			$\alpha_{\rm L}$ & $2.1$ & $\alpha_{\rm N}$ & $4$\\
			$m_{\rm L}$ & $3$ & $m_{\rm N}$ & $1$\\
			\hline
			TBS Parameters & \\
			\hline
			$\lambda_{\rm T}$ & $10$ ${\rm km}^{-2}$ & $\alpha_{\rm T}$ & $4$\\
			\hline
			Environment Parameters~\cite{6863654} &  & &\\
			\hline
			$a$ & $25.27$ & $b$ & $0.5$ \\
			\hline
		\end{tabular}
	\end{table}
	In Fig.~\ref{fig:2}, we evaluate the impact of the charging time, the density of the charging stations, and the battery size on the coverage probability. We select a wide range of values for $T_{\rm ch}$, where low values (such as 5 minutes) represent the scenario of efficient battery swapping while high values (such as 40 minutes) reflect the scenario of slow battery recharging. As can be observed from Fig.~\ref{fig:2}, the charging time has a significant impact on the value of the density of charging stations required to achieve a specific value of coverage probability. For instance, the value of $P_{\rm cov}$ achieved with 1 charging station/km$^2$ with charging time of $40$ minutes, can be achieved with 100 times less density of charging stations if we can provide more efficient charging stations that have a charging time of $5$ minutes. Similar comments also hold for the influence of the battery size $B_{\rm max}$ on the the value of the density of charging stations required to achieve a given value of $P_{\rm cov}$.
	
	In Fig.~\ref{fig:3}, we study the impact of different values of battery size and charging time on the CCDF of ${P_{\rm cov|R_s}}$, $\bar{F}_{P_{\rm cov|R_s}}(X)=1-{F}_{P_{\rm cov|R_s}}(X)$. Agreeing with our comments in Remark~\ref{rem:cov}, we observe that the maximum achievable value of $P_{\rm cov|R_s}$ significantly decrease when we increase the charging time from $5$ to $40$ minutes. 
	
	\section{Conclusion}
	
	In this paper, we derived the coverage probability for a UAV-assisted cellular network as a function of the battery size, the density of the UAV charging stations, and the charging time. Using numerical results, we showed the high impact of the aforementioned system parameters on the system performance. One of the main drawn insights from this paper is the trade-off between deploying high density of low quality charging stations (high charging time) and deploying low density of high quality charging stations (low charging time). Our results showed that we could achieve similar coverage probability with lower density of charging stations if we can reduce the charging time.
	\appendix
	\subsection{Proof of Lemma~\ref{lem:2}}\label{app:2}
	\begin{align}
	\label{2}
	&F_{P_{(\rm a|R_{\rm s})}}(x)=\mathbb{P} ( P_{(\rm a|R_{\rm s})} \leq x ) \nonumber\\
	&=\mathbb{P} \left(\frac{{B}_{\rm max}V-2{P}_{\rm m}{R}_{\rm s}}{{B}_{\rm max}V-2{P}_{\rm m}{R}_{\rm s}+T_{\rm ch}P_{s}V+2{R}_{\rm s}P_{s}} \leq x \right).
	\end{align}
	Given that $P_{(a|R_s)}$ is a decreasing function of $R_{\rm s}$, the preimage can be obtained as follows
	\begin{align}
	F_{P_{(\rm a|R_{\rm s})}}(x)=\mathbb{P} \left(R_{\rm s} \geq \frac{V (B_{\rm max} (x-1)+P_{\rm s} T_{\rm ch} x)}{2 (P_{\rm m} (x-1)-P_{\rm s} x)} \right).
	\end{align}
	
	Given that the minimum value of $R_{\rm s}=0$ and its maximum value for a non-zero availability probability is $\frac{B_{\rm max}V}{2P_{\rm m}}$ then
	$$
	0\leq x \leq \frac{B_{\rm max}}{P_{\rm s}T_{\rm ch}+B_{\rm max}}.
	$$
	
	\subsection{Proof of Lemma~\ref{lem:4}}\label{app:3}
	Recalling that $p_{\rm L}=\eta_{\rm L}\rho_{\rm u}G_{\rm L}R_{\rm U}^{-\alpha_{\rm L}}$ and $p_{\rm N}=\eta_{\rm N}\rho_{\rm u}G_{\rm N}R_{\rm U}^{-\alpha_{\rm N}}$, $P_{\rm cov,U}$ in (\ref{8}) can be rewritten as 
	\begin{small}
		\begin{align}
		\mathbb{E}_{R_{\rm U}}&\left[\mathbb{P}\left(\frac{p_{\rm L}}{\sigma^{2}}\geq\beta|R_{\rm U}\right)P_{\rm L}(R_{\rm U})+\mathbb{P}\left(\frac{p_{\rm N}}{\sigma^{2}}\geq\beta|R_{\rm U}\right)P_{\rm N}(R_{\rm U})\right] \nonumber\\
		=&\mathbb{E}_{R_{\rm U}}\left[\mathbb{P}\left(\frac{p_{\rm L}}{\sigma^{2}}\geq\beta|R_{\rm U}\right)P_{\rm L}(R_{\rm U})\right]\nonumber\\
		&+\mathbb{E}_{R_{\rm U}}\left[\mathbb{P}\left(\frac{p_{\rm N}}{\sigma^{2}}\geq\beta|R_{\rm U}\right)P_{\rm N}(R_{\rm U})\right].
		\end{align}
	\end{small}
	Let 
	\begin{align}
	\label{12}
	P_{\rm cov_{L}}&=\mathbb{E}_{R_{\rm U}}\left[\mathbb{P}\left(\frac{p_{\rm L}}{\sigma^{2}}\geq\beta|R_{\rm U}\right)P_{\rm L}(R_{\rm U})\right],\\
	P_{\rm cov_{N}}&=\mathbb{E}_{R_{\rm U}}\left[\mathbb{P}\left(\frac{p_{\rm N}}{\sigma^{2}}\geq\beta|R_{\rm U}\right)P_{\rm N}(R_{\rm U})\right].
	\end{align}
	Then,
	\begin{align}
	P_{\rm cov_{L}} =&\mathbb{E}_{R_{\rm U}}\left[\mathbb{P}\left(\frac{\eta_{\rm L}\rho_{\rm u}G_{\rm L}R_{\rm U}^{-\alpha_{\rm L}}}{\sigma^{2}}\geq\beta|R_{\rm U} \right)P_{\rm L}(R_{\rm U})\right] \nonumber\\
	\stackrel{(a)}{=}&\int^{\sqrt{h^2+r_{c}^2}}_{h}P_{\rm L}(r)\mathbb{P}(G_{\rm L}\geq g_{l}(r))\frac{2r}{r^{2}_{c}}{\rm d}r \nonumber\\
	\stackrel{(b)}{=}&\int^{\sqrt{h^2+r_{c}^2}}_{h}P_{\rm L}(r)\frac{\Gamma_{u}(m_{\rm L},m_{\rm L}g_{l}(r))}{\Gamma(m_{\rm L})}\frac{2r}{r^{2}_{c}}{\rm d}r \nonumber\\
	\stackrel{(c)}{=}&\int^{\sqrt{h^2+r_{c}^2}}_{h}P_{\rm L}(r) \nonumber\\
	&\times e^{-m_{\rm L}g_{l}(r)}\sum_{k=0}^{m_{\rm L}-1}\frac{(m_{\rm L}g_{l}(r))^{k}}{k!}\frac{2r}{r^{2}_{c}}{\rm d}r.
	\end{align}
	Step (a) is due to the uniform distribution of the users in the disk with radius $r_c$ and $g_{l}(r)=\frac{\beta \sigma^2}{\eta_{\rm L} r^{-\alpha_{\rm L}}\rho_{\rm u}}$, step (b) follows from the definition: $\Bar{F}_{\rm G}(g)=\frac{\Gamma_{u}(m,g)}{\Gamma(m)}$, where $\Gamma_{u}(m,g)=\int^{\infty}_{mg}t^{m-1}e^{-t}dt$ is the upper incomplete Gamma function, and step (c) is form the definition $\frac{\Gamma_{u}(m,g)}{\Gamma(m)}=\exp(-g)\sum^{m-1}_{k=0}\frac{g^{k}}{k!}$.
	
	$P_{\rm cov_{N}}$ can be derived by following similar steps as $P_{\rm cov_{L}}$, therefore omitted here.
	\begin{align}
	P_{\rm cov,T} &=\mathbb{P}(\rho_t HR_{\rm T}^{-\alpha_{\rm T}}\geq\beta) \nonumber\\
	&=\int_{0}^{\infty}\mathbb{P}(H\geq g_{m}(r))f_{R_{\rm T}}(r){\rm d}r \nonumber\\
	&{=}\int_{0}^{\infty}2 \pi r \lambda_{\rm T} e^{\pi \lambda_{\rm T}r^{2}}e^{-g_{m}(r)}{\rm d}r.
	\end{align}
	where $f_{R_{\rm T}}(r)=2\lambda_{\rm T}\pi r\exp(-\lambda_{\rm T}\pi r^2)$ is the contact distance distribution of PPP.
	
	\bibliographystyle{IEEEtran}
	\bibliography{Draft_v0.4.bbl}

\begin{thebibliography}{10}
\providecommand{\url}[1]{#1}
\csname url@samestyle\endcsname
\providecommand{\newblock}{\relax}
\providecommand{\bibinfo}[2]{#2}
\providecommand{\BIBentrySTDinterwordspacing}{\spaceskip=0pt\relax}
\providecommand{\BIBentryALTinterwordstretchfactor}{4}
\providecommand{\BIBentryALTinterwordspacing}{\spaceskip=\fontdimen2\font plus
\BIBentryALTinterwordstretchfactor\fontdimen3\font minus
  \fontdimen4\font\relax}
\providecommand{\BIBforeignlanguage}[2]{{%
\expandafter\ifx\csname l@#1\endcsname\relax
\typeout{** WARNING: IEEEtran.bst: No hyphenation pattern has been}%
\typeout{** loaded for the language `#1'. Using the pattern for}%
\typeout{** the default language instead.}%
\else
\language=\csname l@#1\endcsname
\fi
#2}}
\providecommand{\BIBdecl}{\relax}
\BIBdecl

\bibitem{8713514}
B.~{Galkin}, J.~{Kibilda}, and L.~A. {DaSilva}, ``A stochastic model for {UAV}
  networks positioned above demand hotspots in urban environments,'' \emph{IEEE
  Transactions on Vehicular Technology}, vol.~68, no.~7, pp. 6985--6996, July
  2019.

\bibitem{8660516}
M.~{Mozaffari}, W.~{Saad}, M.~{Bennis}, Y.~{Nam}, and M.~{Debbah}, ``A tutorial
  on {UAVs} for wireless networks: Applications, challenges, and open
  problems,'' \emph{IEEE Communications Surveys Tutorials}, vol.~21, no.~3, pp.
  2334--2360, thirdquarter 2019.

\bibitem{DBLP:journals/corr/abs-1907-04299}
M.~A. Kishk, A.~Bader, and M.-S. Alouini, ``On the {3-D} placement of airborne
  base stations using tethered {UAVs},'' {\em IEEE Transactions on
  Communications}, to appear.

\bibitem{8833522}
M.~{Alzenad} and H.~{Yanikomeroglu}, ``Coverage and rate analysis for vertical
  heterogeneous networks ({VH}et{N}ets),'' \emph{IEEE Transactions on Wireless
  Communications}, vol.~18, no.~12, pp. 5643--5657, Dec. 2019.

\bibitem{8648453}
B.~{Galkin}, J.~{Kibilda}, and L.~A. {DaSilva}, ``{UAVs} as mobile
  infrastructure: Addressing battery lifetime,'' \emph{IEEE Communications
  Magazine}, vol.~57, no.~6, pp. 132--137, June 2019.

\bibitem{8866716}
M.~{Lahmeri}, M.~A. {Kishk}, and M.-S. {Alouini}, ``Stochastic geometry-based
  analysis of airborne base stations with laser-powered {UAVs},'' \emph{IEEE
  Communications Letters}, vol.~24, no.~1, pp. 173--177, Jan. 2020.

\bibitem{kishkmag2019}
M.~A. Kishk, A.~Bader, and M.-S. Alouini, ``Capacity and coverage enhancement
  using long-endurance tethered airborne base stations,'' 2019, available
  online: arxiv.org/abs/1906.11559.

\bibitem{bushnaq2020cellular}
O.~M. Bushnaq, M.~A. Kishk, A.~Celik, M.-S. Alouini, and T.~Y. Al-Naffouri,
  ``Cellular traffic offloading through tethered-{UAV} deployment and user
  association,'' 2020, available online: arxiv.org/abs/2003.00713.

\bibitem{7918510}
M.~{Alzenad}, A.~{El-Keyi}, F.~{Lagum}, and H.~{Yanikomeroglu}, ``{3-D}
  placement of an unmanned aerial vehicle base station {(UAV-BS)} for
  energy-efficient maximal coverage,'' \emph{IEEE Wireless Communications
  Letters}, vol.~6, no.~4, pp. 434--437, Aug. 2017.

\bibitem{7809177}
C.~{Saha}, M.~{Afshang}, and H.~S. {Dhillon}, ``Enriched $k$-tier hetnet model
  to enable the analysis of user-centric small cell deployments,'' \emph{IEEE
  Transactions on Wireless Communications}, vol.~16, no.~3, pp. 1593--1608,
  March 2017.

\bibitem{8663615}
Y.~{Zeng}, J.~{Xu}, and R.~{Zhang}, ``Energy minimization for wireless
  communication with rotary-wing {UAV},'' \emph{IEEE Transactions on Wireless
  Communications}, vol.~18, no.~4, pp. 2329--2345, April 2019.

\bibitem{6863654}
A.~{Al-Hourani}, S.~{Kandeepan}, and S.~{Lardner}, ``Optimal {LAP} altitude for
  maximum coverage,'' \emph{IEEE Wireless Communications Letters}, vol.~3,
  no.~6, pp. 569--572, Dec. 2014.

\end{thebibliography}
	
\end{document}